\documentclass[fleqn,usenatbib]{mnras}
\pdfoutput=1

\usepackage{txfonts}

\usepackage[T1]{fontenc}
\usepackage{ae,aecompl}

\usepackage{graphicx}	
\usepackage{amsmath}	
\usepackage{amssymb}	

\usepackage{relsize}	

\title[Red Giant Noise and Planets]{A simple model to describe intrinsic stellar noise for exoplanet detection around red giants}

\author[T. S. H. North et al.]{
Thomas S. H. North,$^{1,2}$\thanks{E-mail: txn016@bison.ph.bham.ac.uk  (TSHN)}
William J. Chaplin,$^{1,2}$
Ronald L. Gilliland,$^{3,4}$
\newauthor
Daniel Huber,$^{5,6,2}$
Tiago L. Campante,$^{1,2}$
Rasmus Handberg,$^{2,1}$
Mikkel N. Lund,$^{1,2}$
\newauthor
Dimitri Veras,$^{7}$
James S. Kuszlewicz,$^{1,2}$
Will M. Farr$^{1}$,
\\
$^{1}$School of Physics and Astronomy, University of Birmingham, Birmingham, B15 2TT, United Kingdom\\
$^{2}$Stellar Astrophysics Centre (SAC), Department of Physics and Astronomy,Aarhus University,\\ Ny Munkegade 120, DK-8000 Aarhus C, Denmark\\
$^{3}$Department of Astronomy and Astrophysics, and Center for Exoplanets and Habitable Worlds,\\The Pennsylvania State University, 525 Davey Lab, University Park, PA 16802, USA\\
$^{4}$Space Telescope Science Institute, 3700 San Martin Dr., Baltimore, MD 21218, USA\\
$^{5}$Sydney Institute for Astronomy (SIfA), School of Physics, University of Sydney, NSW 2006, Australia\\
$^{6}$SETI Institute, 189 Bernardo Avenue, Mountain View, CA 94043, USA\\
$^{7}$Department of Physics, University of Warwick, Coventry, CV4 7AL, United Kingdom\\
}

\date{Accepted XXX. Received YYY; in original form ZZZ}

\pubyear{2016}

\begin{document}
\label{firstpage}
\pagerange{\pageref{firstpage}--\pageref{lastpage}}
\maketitle

\begin{abstract}

In spite of the huge advances in exoplanet research provided by the
NASA \emph{Kepler} Mission, there remain only a small number of
transit detections around evolved stars. Here we present a
reformulation of the noise properties of red-giant stars, where the
intrinsic stellar granulation, and the stellar oscillations described
by asteroseismology play a key role. The new noise model is a
significant improvement on the current \emph{Kepler} results for
evolved stars. Our noise model may be used to help understand
planet detection thresholds for the ongoing K2 and upcoming TESS
missions, and serve as a predictor of stellar noise for these missions. As an application of our noise model, we explore the minimum detectable planet radii for red giant stars, and find that Neptune sized planets should be detectable around low luminosity red giant branch stars.

\end{abstract}

\begin{keywords}
asteroseismology -- techniques: photometric -- planetary systems
\end{keywords}



\section{Introduction}

Red giants, stars near the end of their life -- which have exhausted
fuseable hydrogen in the stellar core, and bloated massively compared
to their main-sequence radii -- are a relatively new focus for
photometric exoplanet research. The four years of near continuous,
high-quality photometry from the NASA \emph{Kepler} Mission has been a
key driver in studies of exoplanets, including close in planets around evolved stars \citep{Kepler56, TTV, 2014Lillo, Kepler432, Kepler91, 2015Quinn432}. Previous exoplanet
searches around giant stars have primarily been conducted using radial
velocity measurements \citep{Johnson2008, Reffert2014, 2015Quinn}.

One reason for the interest in red giants is that when the Sun reaches
this stage of evolution the fate of the Earth is a contentious matter,
with the ultimate balance between mass loss and the maximum extent of
the Sun being the deciding factors \citep{earthdeath}, along with the influence of tidal decay on the orbit. The timescales
for dynamic evolution of the system are accelerated as the star
evolves, with evidence of several planet hosts on course to devour
their planets \citep{2012engulf}; an example is Kepler-56, a red giant with
two detected transiting planets that are predicted to be consumed by
their star in around 150 million years \citep{Li2014}. 

\emph{Kepler} has provided high precision measurements of stellar
variability, and a host of related phenomena, such as activity,
stellar rotation \citep{mcq2014} and the detection of intrinsic,
oscillations in stars. The analysis of the detected oscillations --
the field of asteroseismology -- in principle provides very precise
constraints on stellar properties, a key ingredient in the
characterisation of exoplanets \citep{Vincent2014}. \emph{Kepler} has observed solar-like oscillations in over 15,000 red giants
\citep{2011Hekk, Mosser2012, 2012A&A...540A.143M,Stello2013}, another reason
that a search for planets around giants is of interest.  Asteroseismology may be used to discriminate
between stars either ascending the red giant branch (RGB), or in the
Helium core burning ``red clump'' (RC) phase \citep{2011Bedding}. This
is particularly important for the possible detection, and existence,
of close-in planets. Asteroseismic results on the stellar angle of inclination of the host star can also
reveal if it is a misaligned system, where the stellar spin axis
and plane of planetary orbits are not coplanar \citep{Kepler56}. Finally, asteroseismology also provides
well-constrained stellar ages \citep{Silva2015}, allowing star and planet formation to
be probed across Galactic history \citep{Kepler444}.

The ability to detect a planetary transit is limited by multiple
factors, the primary factor being the depth of the transit, which is
directly related to the relative size of planet and host star. Another
more subtle issue is the noise properties of the host star, which in
cool main-sequence, sub-giant and red-giant stars can contain
contributions from various stellar signals indicative of granulation,
oscillations and activity. Additionally, there is a shot noise
contribution to be considered and instrumental artefacts. Detecting
the transit signal requires an understanding of the expected noise
properties and the expected appearance of the transit in the
lightcurve.

In this paper we present a simple model of the noise properties
relevant to transit detection around red giants, which employs scaling
relations based on global asteroseismic parameters. The dominant
contributions are those due to granulation and solar-like
oscillations. This model is then used to estimate minimum
    detectable planet radii for different assumed orbital periods.

Readers unfamiliar with asteroseismology will find an introduction to
the relevant parameters in Section \ref{sec:Astero_intro}. The
relevant parameters for the noise model are introduced in Section
\ref{subsec:osc}, and the current \emph{Kepler} noise properties are
discussed in Section \ref{Kepcdpp_sec}. Finally, Section
    \ref{sec:Noise} covers the construction of the noise model and
    discusses the implications of the resulting predictions for
    detecting planets around red giants in \emph{Kepler} data.

\section{Frequency spectrum of red giants: oscillations and granulation}

The noise model detailed below is based on observed stellar
parameters. Given the close connection between stellar granulation and
oscillations, where possible the individual parameters of the model
for the oscillation and granulation components are described in terms
of the asteroseismic parameters, along with additional fundamental stellar
parameters, where appropriate.

We begin here by introducing the relevant asteroseismic parameters,
and the intrinsic stellar properties they relate to. Those already
familiar with asteroseismic parameters can skip to
Section~\ref{subsec:osc}.

\subsection{Asteroseismic global parameters}\label{sec:Astero_intro}

\begin{figure}
	\includegraphics[width=\columnwidth]{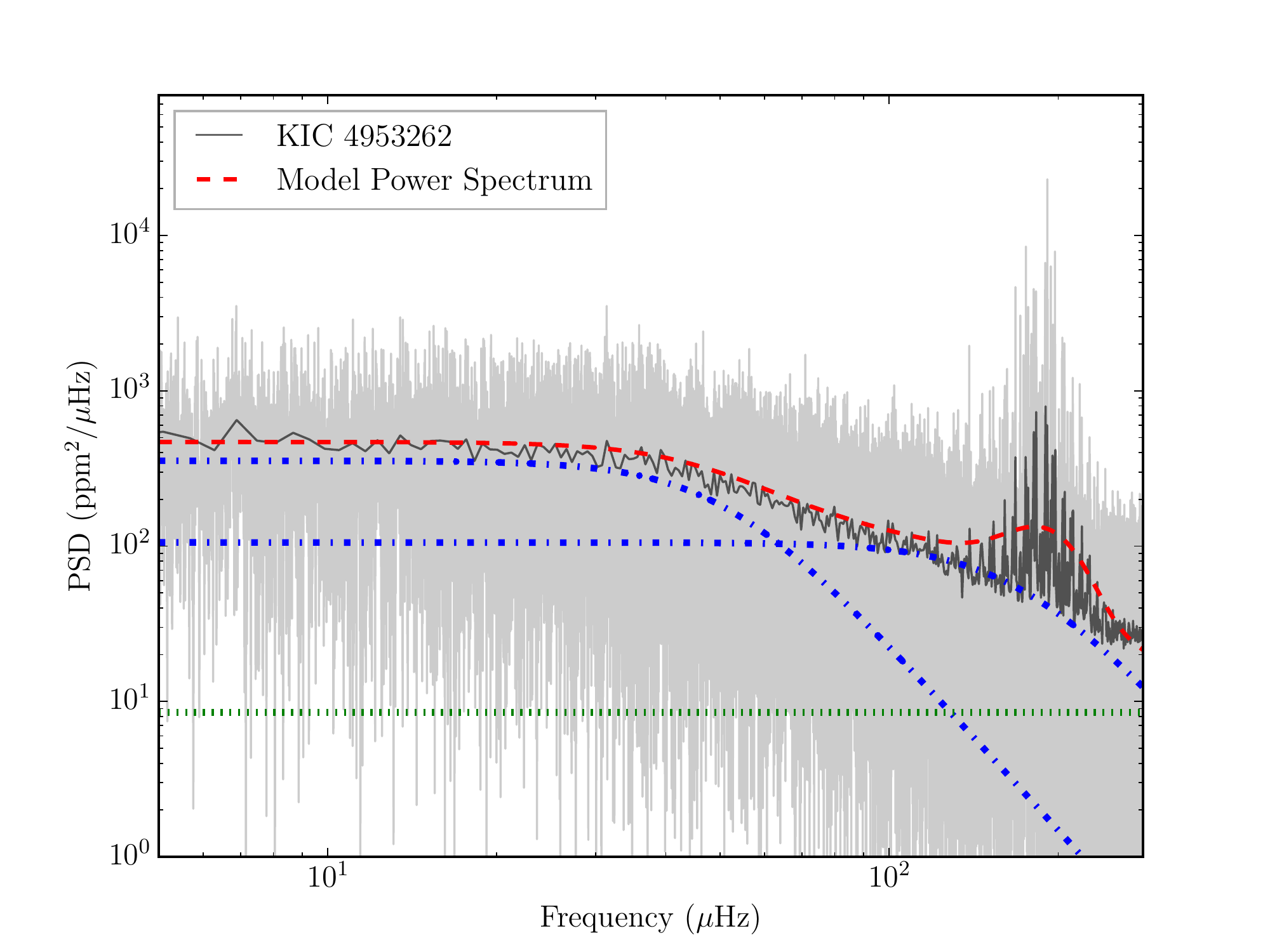}

    \caption{The power density spectrum for KIC 4953262, with the raw
      and smoothed power spectra in grey and black respectively. Green (dotted)
      indicates the shot noise level, showing it is a small factor for
      this star, whilst the blue (dashed dotted) show the two granulation components,
      red (dashed) is total model power spectrum including an oscillation
      component, where the individual modes are not modelled in this
      formulation.}

    \label{fig:psdexample}
\end{figure}

Solar-like oscillations are driven and damped by turbulent convection
in the outer envelope of the star, with the amplitudes of these
signals greatly enhanced in evolved stars \citep{2011Baudin}. Figure \ref{fig:psdexample}
shows an example red-giant frequency power spectrum, made from \emph{Kepler}
data on the target KIC\,4953262. The two main features of the power
spectrum  are the stellar granulation background, and
solar-like oscillations. The oscillations are clearly visible above the
background around 200$\mu$Hz. Additionally, model fits to the
components are overplotted, and will be returned to in Section
\ref{subsec:osc}. For the noise model detailed in Section
\ref{sec:Noise}, the individual oscillation modes do not need to be
modelled, only the oscillation power envelope that contains them.

\begin{figure}
	\includegraphics[width=\columnwidth]{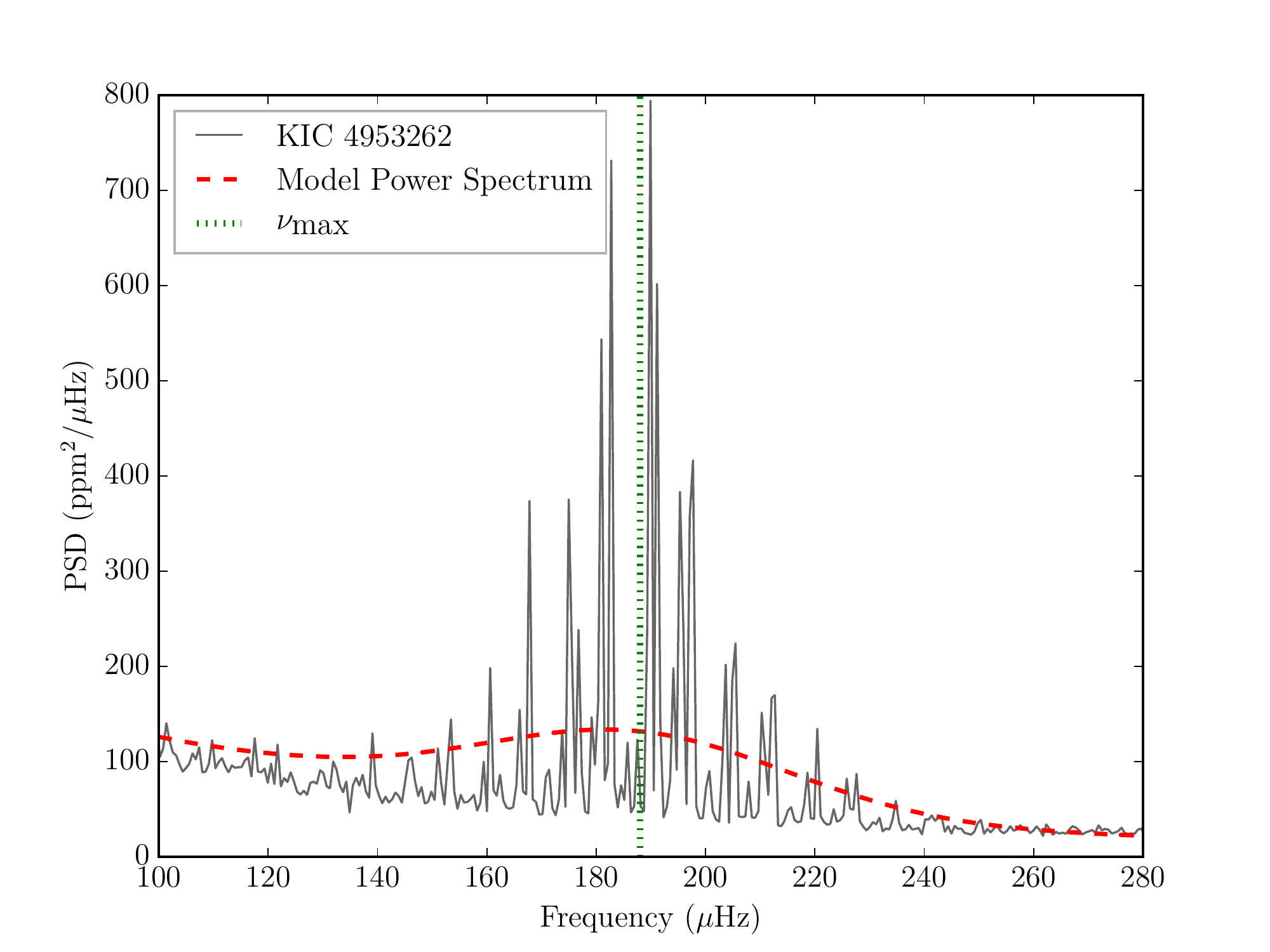}

    \caption{Smoothed power spectrum for KIC 4953262, a known
      oscillating red giant. The vertical dotted line indicates
      $\nu_{\text{max}}$ for this star. Shown in red is a model of the
      power envelope of the oscillation spectrum.}

    \label{fig:linear}
\end{figure}

Figure \ref{fig:linear} shows a zoom of the same power spectrum,
around the region where the detected stellar oscillations are most
prominent. The oscillations appear as fairly evenly spaced peaks in
frequency. Overtones of the same angular degree, $l$, are spaced by
the large frequency separation. The average large separation,
$\Delta\nu$, scales to good approximation with the square-root of mean stellar
density \citep{1986ApJ...306L..37U}, i.e.,
\begin{equation}
\label{delnu}
\frac{\Delta\nu}{\Delta\nu_{\odot}} \simeq \left(\frac{M}{\text{M}_{\odot}}\right)^{0.5}\left(\frac{R}{\text{R}_{\odot}}\right)^{-1.5}.
\end{equation}

The observed power of the mode peaks is modulated by an envelope that
is usually taken as being a Gaussian, centered on the frequency
$\nu_{\text{max}}$, i.e., the frequency at which the detected
oscillations show their strongest amplitudes. This characteristic
frequency can be predicted from fundamental parameters. Its physical
meaning is still debated \citep{2011Belk}, but it scales to very good
approximation with the (isothermal) acoustic cut-off frequency in the
stellar atmosphere, with numerous studies showing
\begin{equation}
\nu_{\text{ac}}\propto\nu_{\text{max}}\propto\frac{c}{H}.
\label{eq:numax1}
\end{equation}
Here, the speed of sound $c\propto \sqrt{T}$, $T$ being the mean local
atmospheric temperature, and $H \propto T/g$ is the pressure scale
height of the atmosphere
\citep{1991Brown,Bedding95}. Equation~\ref{eq:numax1} suggests the use
of a relation scaled to solar values of the form
\begin{equation}
\label{numax}
\frac{\nu_{\text{max}}}{\nu_{\text{max},\odot}} \simeq \frac{g}{g_{\odot}}\left(\frac{T_{\text{eff}}}{\text{T}_{\text{eff},\odot}}\right)^{-1/2},
\end{equation}
where, since oscillations are observed in the stellar photosphere, the
temperature is set to $T=T_{\text{eff}}$.  In this work, the solar
values adopted are: $g_{\odot}=27400\text{cms}^{-2}$,
$\nu_{\text{max},\odot}=3090\mu$Hz and $T_{\text{eff},\odot}=5777$K
\citep{Chaplin2014a}.


Since all the stars considered in this work either have detected
oscillations (real cohort) or would be predicted to show detected
oscillations (synthetic cohort), $\nu_{\textrm{max}}$ will typically
be the parameter we choose to plot against when considering the noise
properties of the stars.

First-order estimates of stellar mass and radius can be 
estimated using the above scaling relations. Combining and
re-arranging Equation \ref{delnu} and Equation \ref{numax} gives \citep{Chaplin2013}
 \begin{equation}
 \frac{M}{\text{M}_{\odot}}=\left(\frac{\nu_{\text{max}}}{\nu_{\text{max},\odot}}\right)^{3} \left(\frac{\Delta\nu}{\Delta\nu_{\odot}}\right)^{-4} \left(\frac{T_{\text{eff}}}{\text{T}_{\text{eff},\odot}}\right)^{1.5},
 \label{eq:mass}
 \end{equation}
and
\begin{equation}
\frac{R}{\text{R}_{\odot}}=\left(\frac{\nu_{\text{max}}}{\nu_{\text{max},\odot}}\right) \left(\frac{\Delta\nu}{\Delta\nu_{\odot}}\right)^{-2} \left(\frac{T_{\text{eff}}}{\text{T}_{\text{eff},\odot}}\right)^{0.5}.
\label{eq:rad}
\end{equation}

With the basic global asteroseismic parameters defined, we now go on
to explore the noise properties of stars in terms of these
parameters. All noise components will be described up to the Nyquist
frequency of the long-cadence \emph{Kepler} data. The 29.4-minute
cadence leads to a Nyquist frequency of $\nu_{\text{Nyq}} \approx 283\mu$Hz
\citep{Koch2010}.

\subsection{Modelling power due to the oscillations}\label{subsec:osc}

For stars that have $\nu_{\text{max}}\lesssim \nu_{\text{Nyq}}$, the
power contained in the oscillations must be considered a component
of the background signal for transit detection. It is sufficient to describe the
contribution due to the oscillations in terms of a Gaussian of excess
power centred around the frequency $\nu_{\text{max}}$ (Equation
\ref{numax}).  The width of the Gaussian is denoted by $\sigma_{\rm
  env}$, as described by Equation 1 in \cite{Mosser2012}, i.e.,
 \begin{equation}
 \label{eq:sigma}
 \sigma_{\rm env}=\frac{\delta_{\text{env}}}{2\sqrt{2\ln2}},
 \end{equation}
with $\delta_{\text{env}}$ describing the full width at half maximum
(FWHM) of the oscillation envelope. The Gaussian also needs a height
(maximum power spectral density), $H$, to give the final form of the
oscillation envelope signature in the power spectrum:
\begin{equation}
\label{eq:oscpower}
\text{PSD}_{\text{osc}}(\nu)=H\exp\left[\frac{-(\nu-\nu_{\text{max}})^2}{2\sigma_{\rm env}^2}\right].
\end{equation}
The height and envelope width, $H$ and $\delta_{\text{env}}$, may be
described in terms of scaling relations expressed in the parameter
$\nu_{\text{max}}$ \citep{Mosser2012}, i.e.,
\begin{equation}
\begin{aligned}
&\delta_{\text{env}}=0.66(\nu_{\text{max}})^{0.88}\\
&H=2.03\times10^{7} (\nu_{\text{max}})^{-2.38} \text{ [ppm}^{2}\mu\text{Hz}^{-1}].
\end{aligned}
\label{eq:deltaenv}
\end{equation}
As noted above, only the envelope describing the total oscillation
power is considered and modelled. The power contained within
individual modes is not required here. Returning to Figures
\ref{fig:psdexample} and \ref{fig:linear}, this envelope is plotted in
red. With the oscillation contribution described, we move to
describing the granulation parameters.

\subsection{Granulation}\label{subsec:gran}

A consequence of visible surface convection is granulation. As hot
material rises on a plume, it cools at the surface and sinks back
down. The stellar material forms cells, with a plume in the centre of
each cell. Photometric granulation signatures for the Sun were
initially modelled by \cite{Harvey1984} as an exponentially decaying signal in the time domain. This is meant to represent the rapid rise in a convective plume,
then the decay as the material cools.  In photometric measurements
this can be considered as the hotter material being intrinsically
brighter, giving a brief spike in flux, before the material cools at
the top of the plume, and grows dimmer, with the process occurring on
some characteristic timescale.

This exponential in time leads to a Lorentzian when described in the
power spectrum (in the frequency domain), and is known as a Harvey
profile. Given that the exact nature of granulation is unclear, and
that this simple formulation does not always appear to fit the
granulation background well, this has in recent years led to a whole
family of ``Harvey-like'' profiles (e.g., see \citealt{Mathur2011}), with
varying formulations and exponents in the functions used.  An
important consideration for our work here is how granulation
properties vary with stellar evolutionary state (once we have selected
a preferred formulation). Does granulation in red giants exhibit the
same behaviour as granulation observed in the Sun? In \cite{Kall2014},
multiple models of granulation were fitted to power spectra over a
range of stellar evolutionary states in cool stars to investigate updated versions
of the original Harvey relation, including a change of exponent.  Observed power spectra often require
the use of multiple granulation components, operating at different
timescales, whereas the original Harvey model used only a single
component, with an exponent of 2. We adopt a two-component granulation model, (described in \citealt{Kall2014} as Model F) i,e.
 \begin{equation}
 \label{Kall}
 \text{PSD}_{\text{gran}}(\nu)=\mathlarger\sum\limits_{i=1}^2 \frac{\xi_{i}a^{2}_{i}/b_{i}}{1+(\nu/b_{i})^{4}}.
 \end{equation} 
Here $\xi_{i}$ is a normalisation constant equal to $2\sqrt{2}/\pi$
for the model, while $a_{i}$ and $b_{i}$ are the granulation amplitude
and characteristic frequency, respectively, of each granulation
component, which are both dependent on the fundamental properties of
the stars. Since the granulation and stellar oscillations are
both driven by convection, it is perhaps not surprising that the granulation amplitude and frequency can
be described by scaling relations based on asteroseismic
parameters. In this case they are based on the frequency of maximum
power $\nu_{\text{max}}$, i.e., from \cite{Kall2014} we have:
\begin{equation}
  \label{eq:gran}
  \begin{aligned}
    a_{1}&=a_{2} = 3710(\nu_{\text{max}})^{-0.613}(M/\textrm{M}_{\odot})^{-0.26},\\
    b_{1} &= 0.317(\nu_{\text{max}})^{0.97},\\
    b_{2} &= 0.948(\nu_{\text{max}})^{0.992},
  \end{aligned}
\end{equation}
with an additional constraint from the stellar mass for the granulation amplitude (which may be derived from
Equation~\ref{eq:mass}, using $\nu_{\text{max}}$, $\Delta\nu$ and
$T_{\rm eff}$ as input). Whilst in \cite{Kall2014} both amplitude
components ($a_{1}$ and $a_{2}$) were allowed to vary during the
fitting procedure, the final relation produced used a single amplitude
relation for both components.  The mass-dependent formulation was also
found to be a better fit to the real data, and as such is the
formulation used here for the granulation amplitude. For the cohort of
real asteroseismic stars considered below (see
Section~\ref{Kepcdpp_sec}) we estimate stellar masses and radii using
the scaling relations defined in Equations~\ref{eq:mass}
and~\ref{eq:rad}, with the solar value taken to be
135.1$\mu$Hz in this work \citep{Chaplin2014a}

Returning to Figure \ref{fig:psdexample}, the two granulation
parameters plotted in blue, along with the oscillation envelope
detailed above, make up the model power spectrum in red. Additionally,
the shot noise component is plotted in green, clearly a small
contribution in this power spectrum. It is from the model spectrum
that we may compute a suitable noise metric for the star.

\section{\emph{Kepler} CDPP}\label{Kepcdpp_sec}

The primary \emph{Kepler} noise metric is the CDPP, or Combined
Differential Photometric Precision, which is designed to describe the
noise properties of a star centred around a timescale of 6.5\,hr \citep{Chris2012, Gill2011}. This is
half the timescale on which an Earth analogue would transit a Sun-like star. Throughout the paper references to \emph{Kepler} CDPP will refer to the 6.5\,hr timescale. The CDPP will be composed of a shot noise component due to
counting signals, but a significant stellar variability term should
also be present. The nature of the stellar variability is dependent on
the intrinsic stellar properties, with possible contributions from
granulation, oscillations and activity.

\emph{Kepler} lightcurves are produced in the Presearch Data
Conditioning module (PDC) \citep{Jenkins2010a, 2012Smith, 2012Stumpe}, and in general the PDC
pipeline is highly successful at removing systematics and instrumental
effects in the lightcurves. However the PDC also removes real
astrophysical signal at long periods \citep{murphy2014investigating}. This is of interest for evolved
stars, having significant low-frequency signals typical of granulation
and intrinsic oscillations. This loss of real signal has the effect of
artificially reducing the reported CDPP, since real variability has
been removed.

\begin{figure}
	\centering
	\includegraphics[width=\columnwidth]{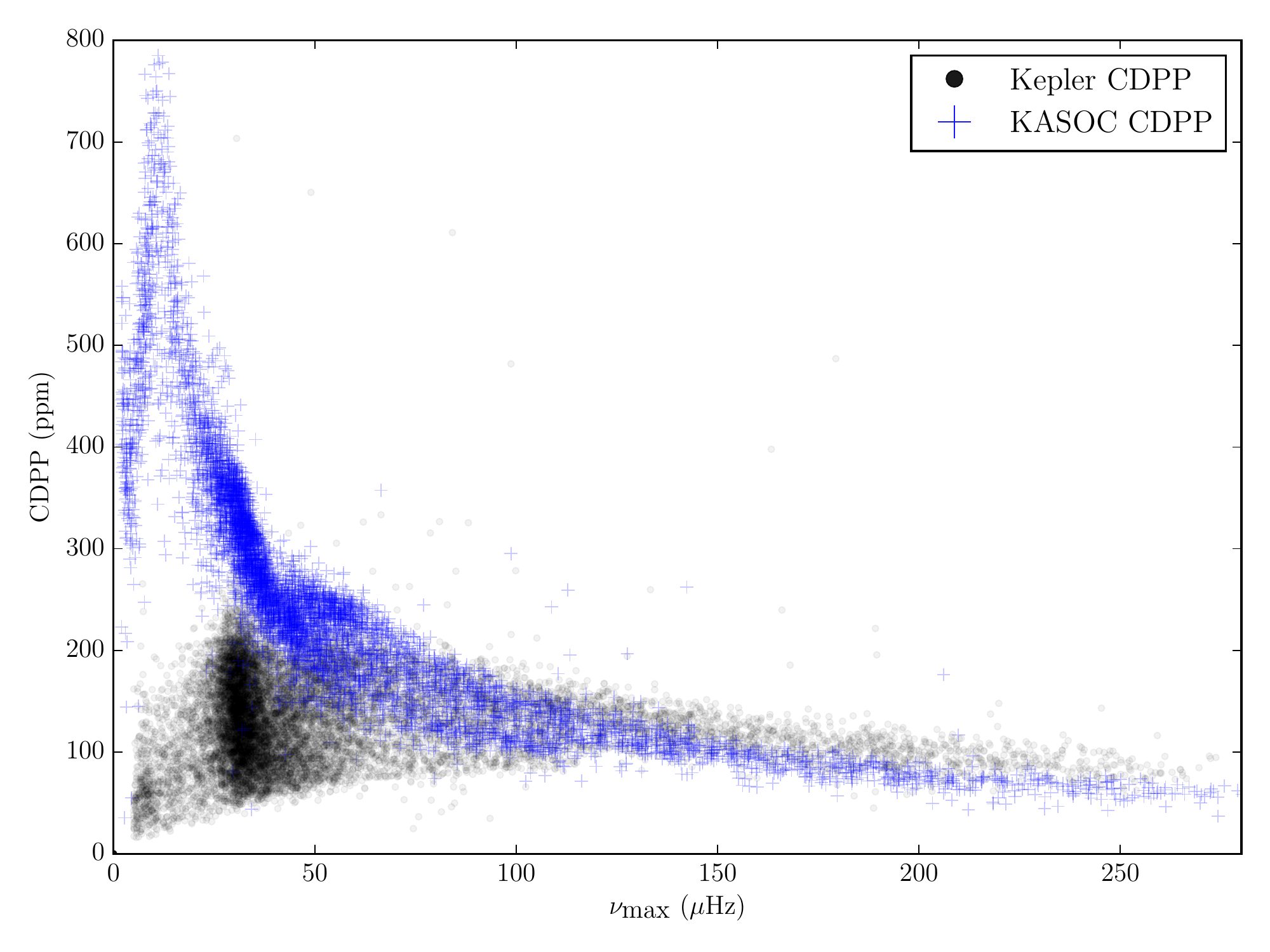}

      \caption{The reported CDPP for 13,000 evolved stars
        \citep{Stello2013}  plotted against the reported
        asteroseismic $\nu_{\text{max}}$ in black. The overall trend with
        decreasing $\nu_{\text{max}}$ is expected due to the
        increasing granulation amplitude (see Equation \ref{eq:gran}),
        but the turnover and spread below $100\mu$Hz is evidence of
        the PDC pipeline removing astrophysical signal. Blue points are the result of work from KASOC (see text).}

    \label{fig:13kcdpp}
\end{figure}

Figure \ref{fig:13kcdpp} shows the reported CDPP for 13,000 red giants
observed by \emph{Kepler}. The reported CDPP appears to show increased
scatter and attenuation at $\nu_{\text{max}}<100\mu$Hz, i.e., in the
more evolved stars in the cohort. The level of signal attenuation was
explored by \cite{2015Gill} and \cite{datarelease21}. Long-period
signals were injected into lightcurves, and attempts made to recover
them after PDC processing. It was found that signals on timescales
longer than a day showed attenuation. The scatter below $100\mu$Hz in
Figure \ref{fig:13kcdpp} suggests that variability on timescales
longer even than only 0.1\,days will suffer some signal
loss. \cite{2015Gill} also note that small-amplitude signals suffer
more attenuation, in relative terms, than large-amplitude signals at
the same frequency (period).

Taken at face value, Figure \ref{fig:13kcdpp} suggests that some of
the low $\nu_{\text{max}}$ (larger, more evolved) stars would be ideal
for planet searches, since they appear to be photometrically
quiet. However the turnover around $100\mu$Hz is unphysical, a
consequence of the PDC lightcurve processing
\citep{datarelease21,Stumpe2014}. This is the primary motivation to
formulate an accurate model of the CDPP for evolved stars.

The data plotted in blue are the CDPP values calculated from
lightcurves produced by an independent processing of the raw
\emph{Kepler} pixel data by the \emph{Kepler} Asteroseismic Science
Operations Center (KASOC) pipeline \citep{Handberg2014}. This
pipeline was intentionally designed to preserve astrophysical signal
on longer timescales, and does not show the same marked attenuation as
the PDC data. As we shall now go on to discuss, our simple noise model
-- which is based on the scaling relations outlined above -- is able
to reproduce the observed KASOC CDPP values.

\section{Noise Model}\label{sec:Noise}

Of the 13,000 stars in Figure \ref{fig:13kcdpp}, $6400$ were
identified as stars ascending the RGB (Elsworth, private comm).  For each of these stars we
constructed basic model power spectra up to the Nyquist frequency of
$283\,\rm \mu Hz$. The granulation and oscillation power envelope
contributions to the spectrum -- which below we label as $P_{\rm g}$
and $P_{\rm o}$ -- were modelled as in Sections~\ref{subsec:osc}
and~\ref{subsec:gran}, using the measured asteroseismic parameters \citep{Stello2013} as
input. The flat shot noise contribution $P_{\rm s}$ was modelled
according to the upper envelope model described in
\cite{Jenkins2010}. The RMS noise per long cadence in the time domain
is
 \begin{equation}
 \sigma_{\text{s}}=\sqrt{c + 7\times10^{7}}/c,
 \end{equation}
where 
 \begin{equation}
 c = 3.46\times10^{0.4\times(12-Kp)+8}
 \end{equation}
is the number of detected electrons per long cadence. The flat
power-spectral density in the frequency domain then corresponds to:
 \begin{equation}
 P_{\textrm{s}}=2\times10^{-6}\sigma_{\textrm{s}}^2\Delta t
 \label{Pshot}
 \end{equation}
where $\Delta t$ is the 29.4-minute cadence.  Components due to the
near-surface magnetic activity were not considered due to the evolved state of these stars. As we shall see below, this assumption
appears to be validated by the good match of our model to the
observations.

The model estimate of the CDPP may then be constructed as follows
 \begin{equation}
 \sigma_{\textrm{CDPP}}=\left(\Delta_{T} \int_{0}^{\nu_{\textrm{Nyq}}}\!F(\nu) \times
 \left[P_{\rm g}+P_{\rm o}+P_{\rm s}\right]\right)^{0.5},
 \label{eq:fullnoise}
  \end{equation}
where $\Delta_T$ is the resolution on which the artificial power
spectra were computed and $F(\nu)$ represents the bandpass filter
response for the model CDPP, which is comprised of high- and low-pass
responses.  As noted in \cite{Gill2011} the high-pass response may be
described by a 2-day Savitsky-Golay filter \citep{Sav1964}, whilst the
low-pass response is a 6.5-hr sinc-squared function. The low-pass response
ensures that the filter has zeros at harmonics of the 6.5-hr Earth-Sun
half-transit duration, so that when constructing the
noise metric transit signal is not included as misidentified stellar
variability. The high-pass filter suppresses the model power spectral
density around zero frequency. The filter has been tested against
\emph{Kepler} stars to ensure that the final values are similar to the
PDC derived CDPP, for stars where no signal attenuation is
occurs.

The attenuation of the signal due to the finite sampling time of \emph{Kepler} is not considered here, due to the negligible influence of the effect around the region of the bandpass filter.

\begin{figure}
	\includegraphics[width=\columnwidth]{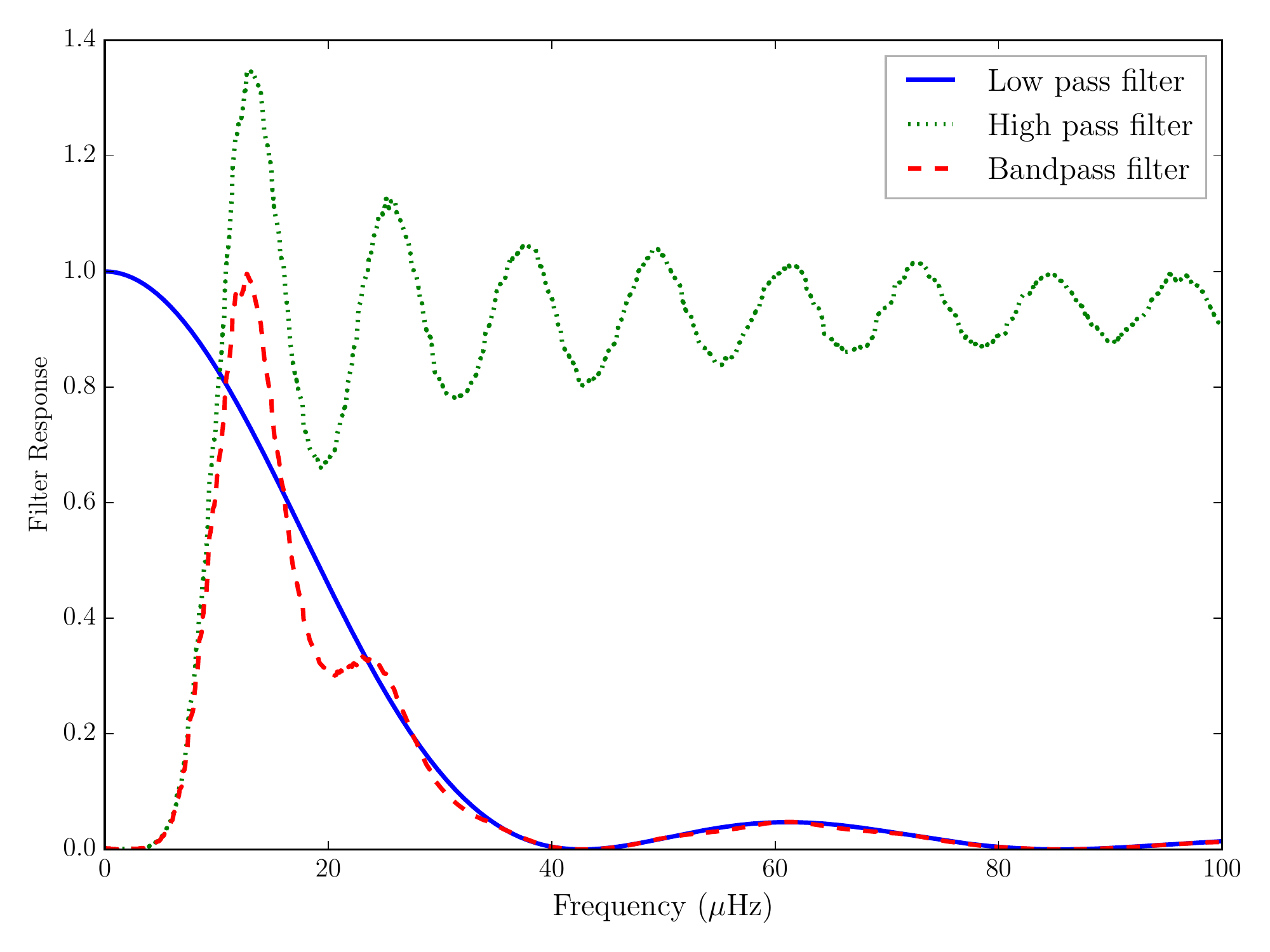}
    \caption{Filter response, with the Savitsky-Golay high-pass in green, 6.5-hr sinc-squared in blue, and the combined filter in red.}
    \label{fig:filter}
\end{figure}

\begin{figure}
	\includegraphics[width=\columnwidth]{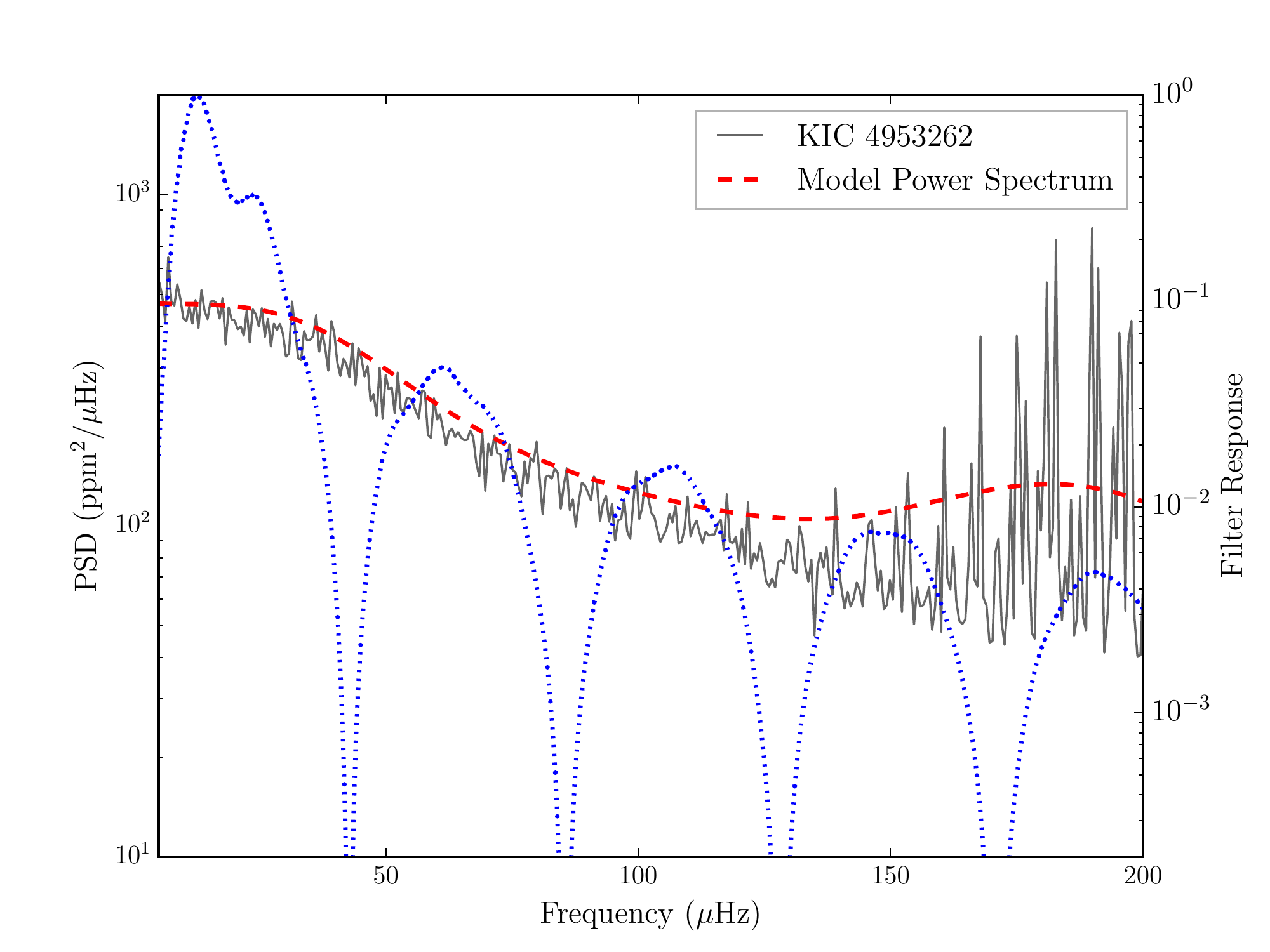}

    \caption{Filter response overplotted on KIC 4953262 power
      spectrum. Clearly most of the signal involved in the
      construction of the noise metric appears in the region
      $0<\nu\lesssim40\mu$Hz. The filter response is shown on
      a log scale to emphasise regions of the power spectrum that
      contribute to the noise metric.}

    \label{fig:filtpsd}
\end{figure}

Figure \ref{fig:filter} shows the main bandpass of the filter, whilst
Figure \ref{fig:filtpsd} shows the filter imposed on a typical red
giant power spectrum to indicate regions of the spectrum captured by
the filter. Since the filter has higher-frequency structure, i.e.,
``ringing", the CDPP of even low-luminosity red giants with
$\nu_{\text{max}}$ values above $200\,\rm \mu Hz$ will have some
contribution from the oscillations. However it should be clear that
for low-luminosity red giants the primary contribution to the stellar
noise will come from the stellar granulation, with the oscillations
being a relatively minor, but not insignificant, contribution.

\begin{figure}
	\includegraphics[width=\columnwidth]{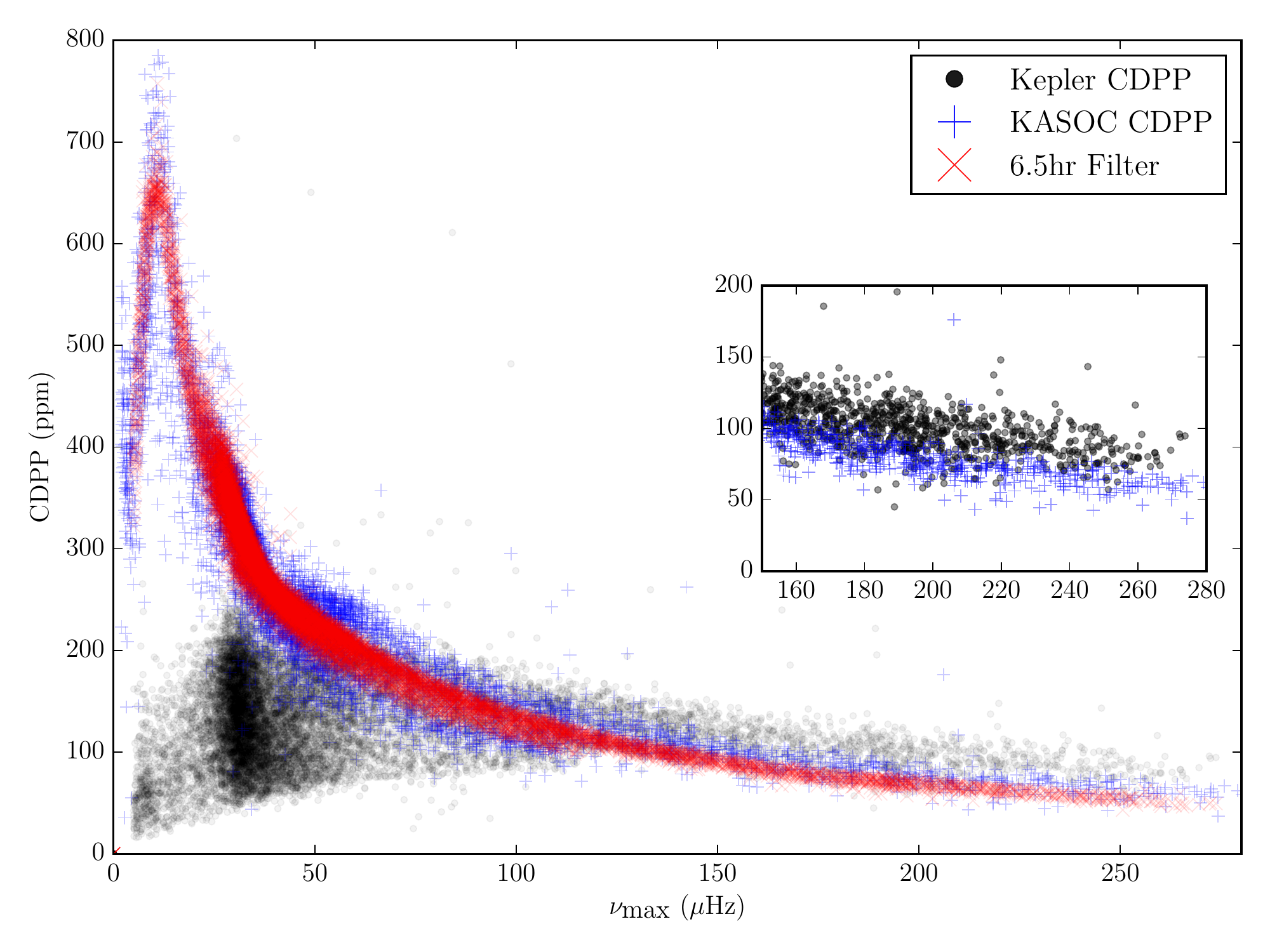}

    \caption{The model CDPP shows a strong trend with
      $\nu_{\text{max}}$. Stars at lower $\nu_{\text{max}}$ represent
      larger stars, with larger granulation signal, since the amplitude
      scales with $\nu_{\text{max}}$ (see Eq \ref{eq:gran}). At low
      frequencies around 10$\mu$Hz, the contribution from the stellar
      oscillations is of the same order as the granulation
      background. The KASOC results are also reproduced and show
      good agreement with the model results. The inset focuses on the high $\nu_{\text{max}}$ ($\nu_{\text{max}}>150\mu$Hz) stars, and shows that the KASOC results show significantly less noise than the PDC derived CDPP. }

    \label{fig:rework}

\end{figure}

Figure \ref{fig:rework} shows the model-estimated CDPP values in red,
overlaid on the observed CDPP values from Figure \ref{fig:13kcdpp},
(PDC pipeline CDPP values in black and the KASOC pipeline CDPP values
in blue). We see good agreement between the model and the observed
KASOC pipeline values.  This is a clear indication that the model used
is sufficiently robust, and additionally that a stellar activity
component is not required for these stars. The turnover
    around 10$\mu$Hz is due to the oscillation envelope passing
    through the frequency bandpass of the filter. The additional
scatter seen in the KASOC results around 50$\mu$Hz is due to the
presence of RC stars, which do not obey the scaling relations used in
construction of power spectra in the same way as stars on the RGB, we
therefore removed these stars in the work that follows.  The clump
stars were also removed due to the assumption that upon ascent up the
RGB, any existing low period planetary system will have been engulfed
by the star. As such they are of little relevance when considering the
potential planet yield left in \emph{Kepler} data.

\begin{figure}
	\includegraphics[width=\columnwidth]{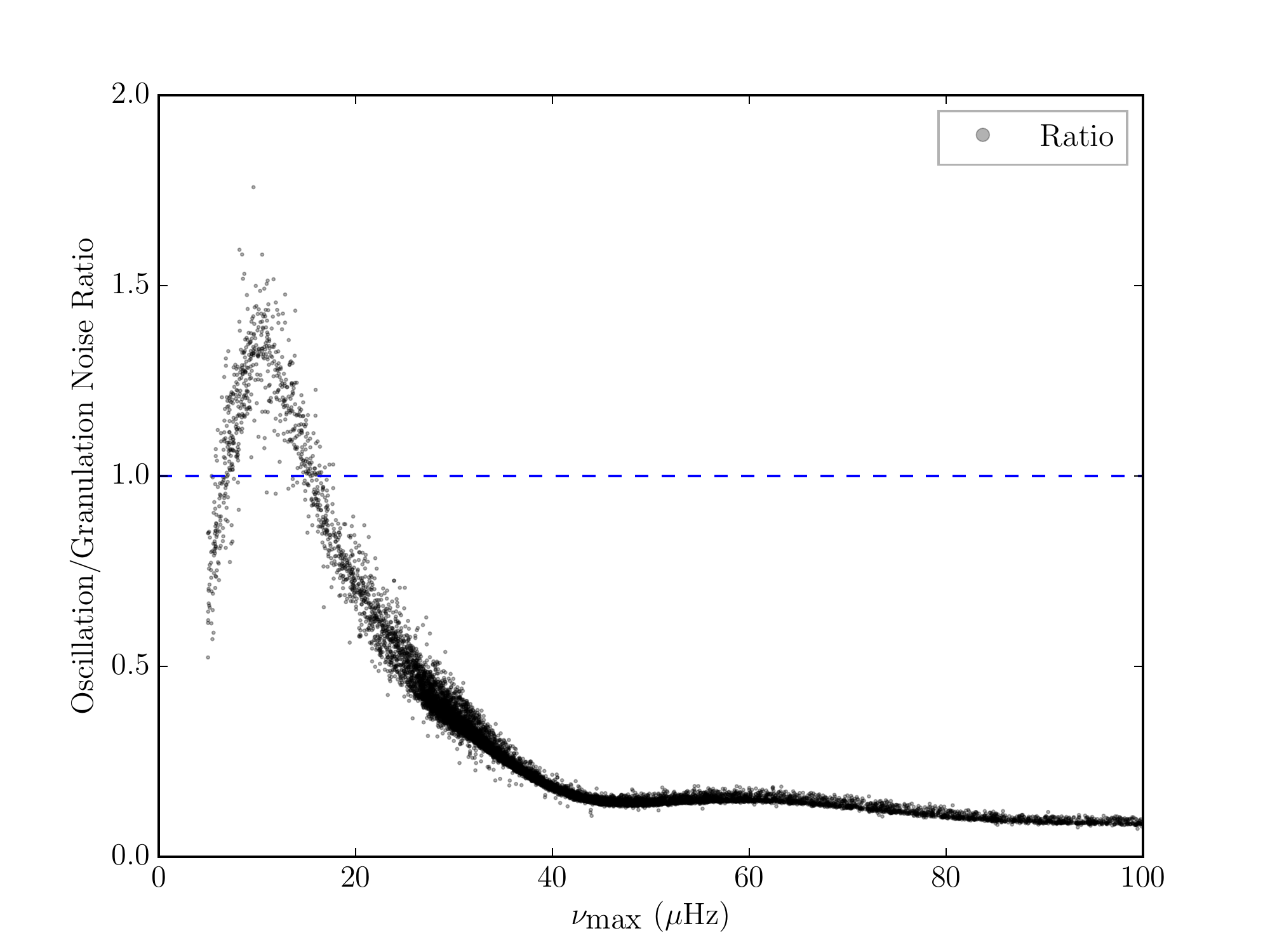}

    \caption{The ratio of the contribution to the model CDPP between
      oscillations and granulation, for 6400 known red giants. The
      dashed line marks unity.}

    \label{fig:granosc}
\end{figure}

Figure \ref{fig:granosc} also demonstrates that the intrinsic stellar
oscillations are a key component of the stellar noise for low
$\nu_{\text{max}}$, high-luminosity RGBs.  In the region around the
turnover ($\nu~10\mu$Hz) in Figure \ref{fig:rework}, the signal from oscillations
dominates by a factor of ${\sim}1.5$; there is also an enhancement in
the oscillation contribution around the first ringing of the filter at
$60\mu$Hz because this is where the oscillation envelope passes
through the filter (with $\nu_{\text{max}}$ aligning with a local
maximum in the filter bandpass). It is important to note that even
when granulation is the dominant noise source, the stellar
oscillations remain a significant factor. 

Finally, it should be noted from Figure \ref{fig:rework} that
low-luminosity giants near the base of the RGB show lower noise in the
KASOC pipeline data than in the PDC data, as highlighted in the inset. This would have consequences
for the detection yield from these stars.

Having established earlier that our model does a good job of
describing the intrinsic stellar noise for evolved stars, we
    go on to apply this CDPP to estimate the minimum detectable planet
    radii around red giant stars in \emph{Kepler} data.

\subsection{Minimum Radius detection}\label{sec:rmin}

The canonical \emph{Kepler} CDPP is designed to capture the noise
properties around a 6.5-hr timescale, related to the transit timescale
of an Earth analogue. But is this filter appropriate to the red-giant
case?  The basic form of the transit duration equation
\citep{Seager2003, Winn2010} is
\begin{equation}
t_{\text{dur}}=\frac{R_{\star}P}{\pi a} \left( 1 - b^2 \right)^{0.5},
\label{eq:tdur_simp}
\end{equation}
where $P$ is the orbital period, $a$ is the semi-major axis, and $b$
is the impact parameter, and there is the implicit assumption of circular orbits. This may be re-written in the form:
\begin{equation}
t_{\text{dur}}=\left[ \frac{4R_{\star}^2a}{GM_{\star}} \left( 1 - b^2
  \right) \right]^{0.5}.
\label{eq:tdur_simp1}
\end{equation}
The maximum transit duration (for $b=0$) is therefore proportional to
$R_{\star}a^{0.5}$. This can potentially vary anywhere from an
Earth-analogue duration (e.g., Kepler-56b, a short-period
planet around another low luminosity red giant, with a transit duration of 13.3hrs) up to durations
exceeding one day (e.g., wide orbits around low-luminosity RGB stars,
or closer orbits around more evolved giants).

Since the range of possible transit durations is so broad for stars
ascending the RGB, the noise properties being considered need to
capture the stellar variability over the relevant timescales. A 6.5-hr
filter turns out to be more appropriate than it might at first
seem. To explain why, we return to Figure \ref{fig:filter}.  The
maximum of the bandpass of the filter is at 12.5$\mu$Hz, a timescale
of around 22\,hours.  The half power points of the bandpass lie at
9.2$\mu$Hz and 16.6$\mu$Hz, corresponding to 30.2 and 16.7\,hours
respectively. There is also a significant contribution to the bandpass
at even shorter periods (i.e., note the secondary peak at around
25$\mu$Hz, which corresponds to about 11\,hours). As we shall see
below, because the chances of detecting planets around very evolved
red giants -- where transit durations would be much longer than a day
-- are so low, our numbers above indicate that the current filter
already does a reasonable job of capturing the necessary timescales of
interest for transits of lower luminosity red giants.

The CDPP values from our model as inputs to calculate a
minimum detectable planet radius for each of the \emph{Kepler} RGB stars, according to
Equation~1 in \cite{Howard2012}:
\begin{equation}
\label{eq:rmin}
R_{\text{min}}=R_{\star}\left(\text{SNR} \times
\sigma_{\text{CDPP}}\right)^{1/2}\left(\frac{6.5\text{hr}}{n_{\text{tr
  }}t_{\text{dur}}}\right)^{1/4},
\end{equation}
The assumed detection signal-to-noise ratio was taken as SNR=10, this value is adopted as a ``secure'' detection threshold. This is stronger than the 7.1$\sigma$ threshold used in the \emph{Kepler} mission for transit detections \citep{Jenkins2010a} to ensure these planets would be detected (see \citealt{Borucki2011, Howard2012, Fressin2013, Chris2013}) .

The transit duration was calculated according to Equation
\ref{eq:tdur_simp}, taking $b=0$; the stellar radius was taken to be
the asteroseismically determined value from Equation \ref{eq:rad}; and
$n$, the number of observed transits, was assumed to equal
$n=4$yr/Period(yr), rounded down to the nearest integer. The factor of
6.5 in Equation \ref{eq:rmin} accounts for the timescale on which the
CDPP is calculated compared to the transit duration. It should also be
noted that the 4-year factor in the number of transits assumes all
stars were observed continuously for the entire duration of the
\emph{Kepler} mission, any missing transits would increase the minimum
detectable radius.

\begin{figure}
\includegraphics[width=\columnwidth]{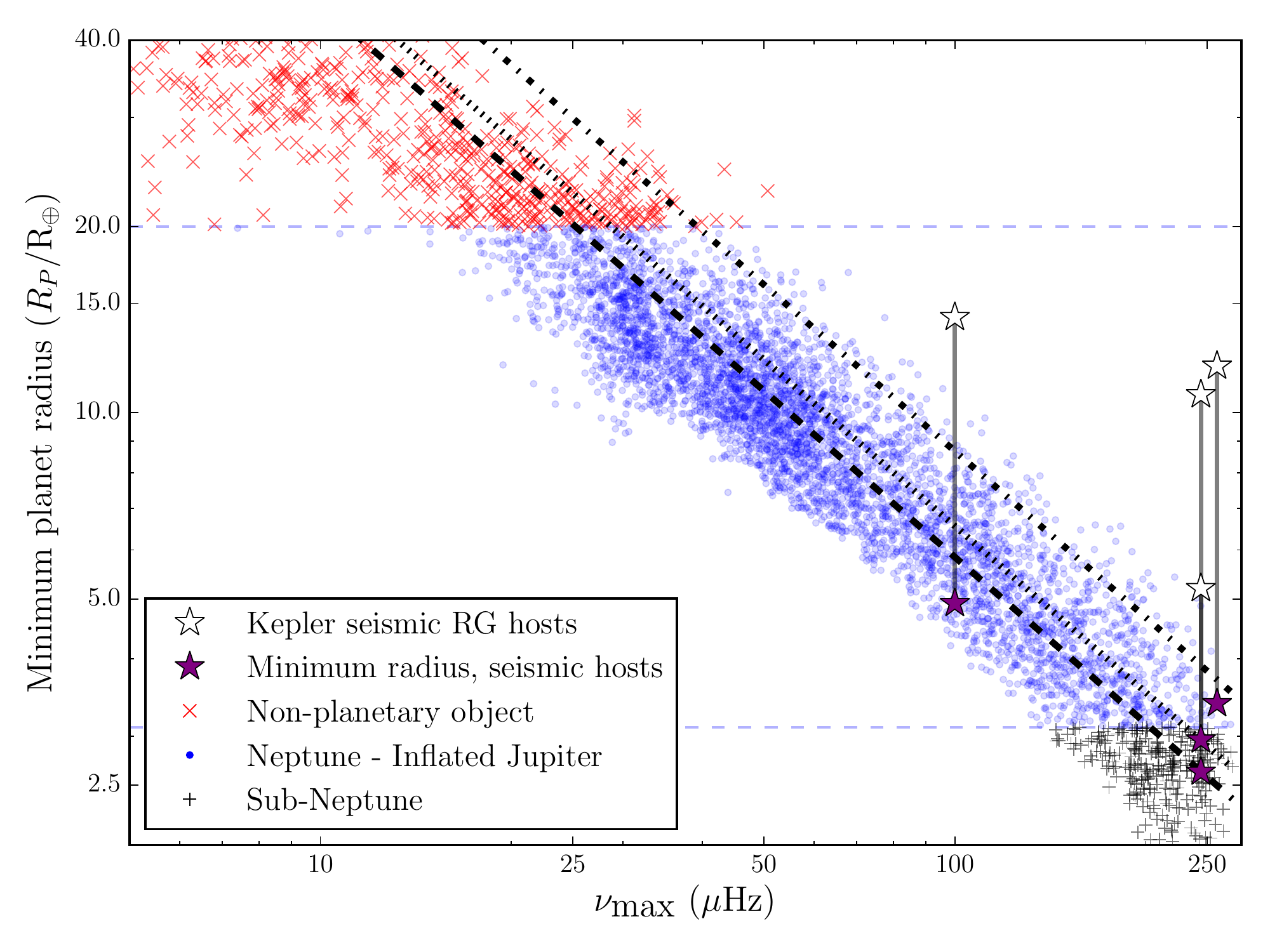}

\caption{Minimum detection radius in Earth radii, for the 6400 \emph{Kepler} stars. Clearly this is a
  strong function of $\nu_{\text{max}}$, in this case a proxy for
  stellar radius. The diagonal lines are fits to a power-law
      relation between $\nu_{\textrm{max}}$ and $R_{\text{min}}$ for
      assumed periods of 10 days (dashed), 20 days (dotted) and 100
      days (dot-dashed). Radii of known planets (open stars) and the
      corresponding estimated minimum radii (filled stars) for the
      same systems are also shown, connected by vertical black
      lines. Points and crosses indicate the minimum radii for illustrative distribution described in the text}

\label{fig:rmin20}
\end{figure}

The diagonal lines in Figure~\ref{fig:rmin20} show power-law
    fits to $\nu_{\text{max}}$ of the calculated minimum detection
    radii $R_{\text{min}}$ of the 6400 \emph{Kepler} stars, assuming fixed
    orbital periods of 10\, days (dashed line), 20\,days (dotted line)
    and 100\,days (dot-dashed line), respectively. The vertical offset
    seen between the diagonal lines is due to the reduced number of
    transits seen for longer period planets. The minimum radii here
    were calculated using the model CDPP predictions. But we could
    also have used the KASOC CDPP data, which give very similar
    results. The true, underlying period distribution for planets
    orbiting evolved hosts is of course very poorly constrained. For
    illustrative purposes only, we have also calculated minimum radii
    using an underlying distribution that is consistent with results
    on confirmed \emph{Kepler} planets, with data taken from the NASA
    Exoplanet Archive
    \citep{2013Akeson} \footnote{http://exoplanetarchive.ipac.caltech.edu/}. These
    data are well described by a log-normal distribution, with the
    underlying normal distribution having a mean and standard
    deviation of 2.47 and 1.23 in $\log_e P$. The results are plotted
    on Figure~\ref{fig:rmin20}, blue dots are super-Earth to Jupiter sized objects, whilst red crosses are objects with minimum radii greater than that which is feasible for a planet. Black crosses indicate a minimum radii of less than the radius of Neptune.

Figure \ref{fig:rmin20} shows that even the most inflated hot-Jupiter
planets will be undetectable around high RGB stars (i.e., stars with
low $\nu_{\textrm{max}}$). This is due to the large radii of
    these stars, and the resulting small transit depths. Due to the
inflated nature of the stars themselves, finding Earth-like planets at
high SNR will most likely prove unfeasible across the entire
population of evolved stars. For low-luminosity red giants, there is
the potential to reach super-Earth sized planets. However it is
apparent that the focus for planets around red-giant hosts should be
Neptune to Jupiter-sized giant planets. 

Radii of known planets (open stars) and the estimated minimum
    radii (filled stars) for the same systems are also shown on Figure
    \ref{fig:rmin20}, connected by vertical black lines.  As can
also be seen, the currently known transiting planets around evolved
hosts sit on the upper edge of the distribution in planet radius and
$\nu_{\text{max}}$.  The lack of detections around
low-$\nu_{\text{max}}$ stars suggests that any systematic search for
planets around evolved hosts should instead concentrate on
low-luminosity RGB stars.  We note that we might expect radii for
actual detections to cover a range of radii at and above the minimum
radii and this is what we see in Figure \ref{fig:rmin20}, albeit for a
very small sample. 


\begin{figure}
\includegraphics[width=\columnwidth]{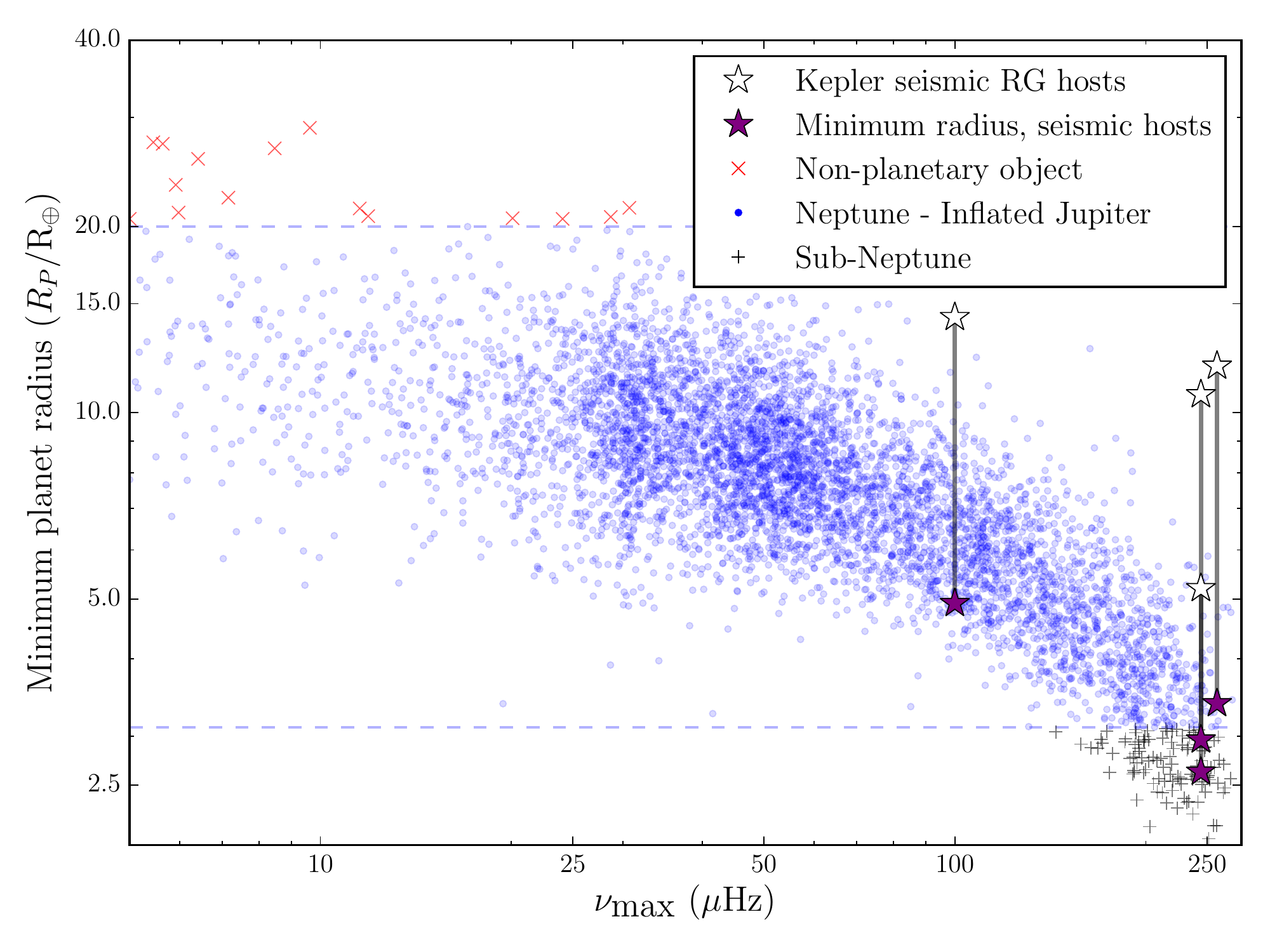}

\caption{Minimum detection radius in Earth radii, if the PDC CDPP results are used. Again the 20 day distribution has been used here. Clearly the PDC results would suggest that planets would be detectable around low $\nu_{\text{max}}$ stars, but this is purely an effect of the PDC processing producing anomalously low  CDPP values.}

\label{fig:rminPDC}
\end{figure}

Figure \ref{fig:rminPDC} shows the same minimum radius calculation
using the current \emph{Kepler} PDC derived CDPP values. These results
would (incorrectly) suggest that planets could be detected around low
$\nu_{\text{max}}$ stars due to the aforementioned attenuation of
intrinsic stellar signals on long timescales. For the high
$\nu_{\text{max}}$ stars, the minimum radii are also larger than the
results for the for updated noise model CDPP described here.


As stars ascend the RGB, planets on short periods are rapidly engulfed
by the expanding star. Additionally the tidal decay timescale
decreases for evolved stars \citep{2013Schlaufman}, e.g., the
Kepler-56 system, where the planets are likely to be engulfed within
${\sim}$100\,Myr \citep{Li2014}. Even without consideration
    of tidal decay, for the case of evolved RGB hosts, planets on
    short periods, and many cases in the \emph{Kepler} period
    distribution described above, would have to exist inside the
    stellar envelope (these cases have been removed from Figure
    \ref{fig:rmin20}).

\subsection{Transit Injection Test}
\begin{figure}
\includegraphics[width=\columnwidth]{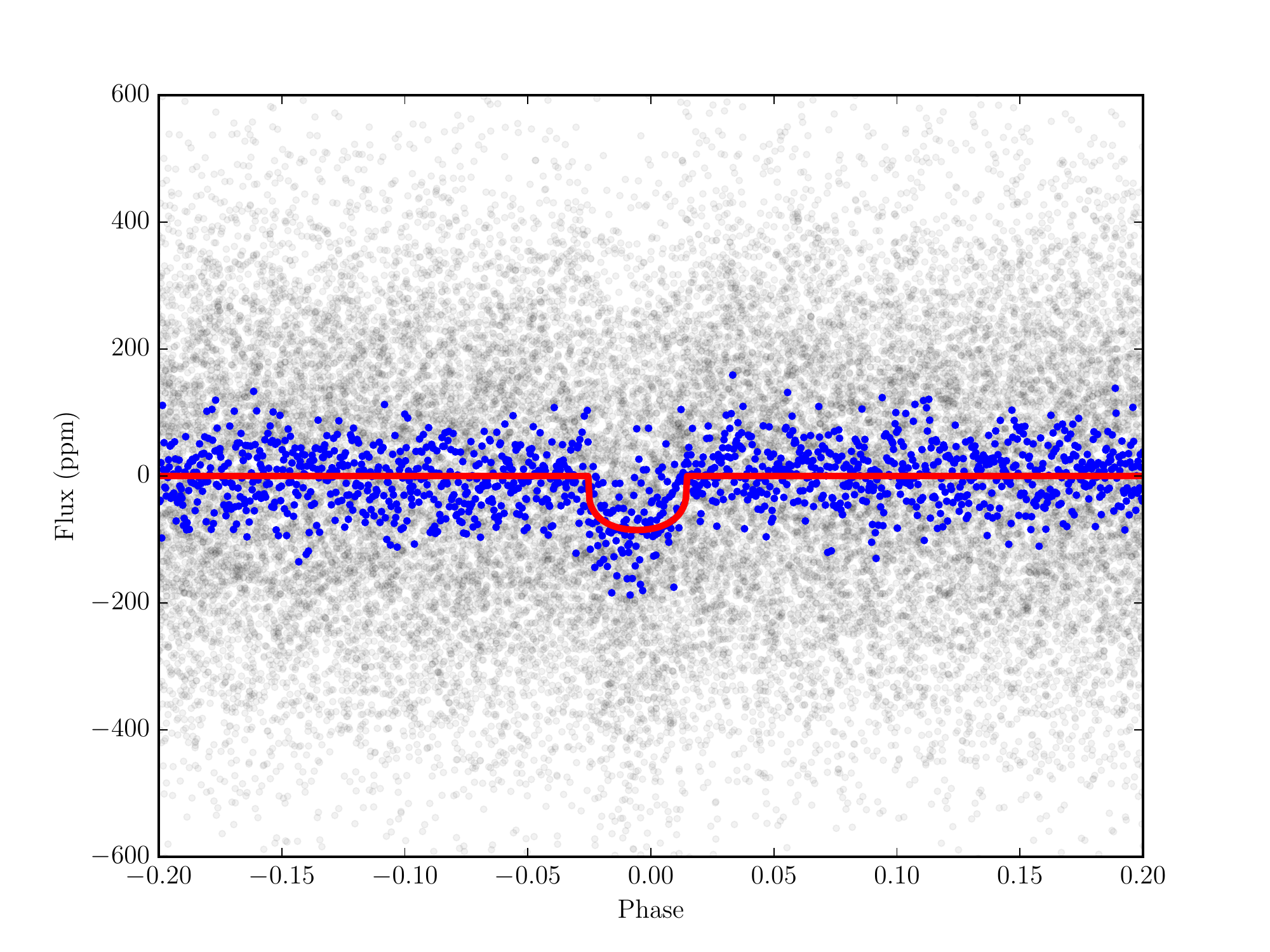}

\caption{Injected transit into \emph{Kepler} detrended lightcurve, folded on the 20 day period of injected planet (black). Also plotted is binned lightcurve, folded on period (blue), along with model for the planet injection (red)}

\label{fig:folded}
\end{figure}

To ensure the results for the minimum detection radius in Figure
\ref{fig:rmin20} are reasonable, a sensible test was to inject transit
signals into real \emph{Kepler} data and attempt to recover the
transit signal. As an example a red giant with similar stellar and
asteroseismic properties ($\nu_{\text{max}}=255\mu$Hz) to Kepler-56,
but with no known transits, was selected and a transit signal injected
into the detrended lightcurve. A planet with the minimum detection
radius ($R_{\textrm{min}}=2.25R_{\oplus}$, for a planet on a 20 day
orbit, at SNR=10) was injected into the lightcurve on a 20 day orbit,
and was recovered using a box-least squares algorithm\footnote{python
  implementation of BLS created by Dan Foreman-Mackey and Ruth Angus
  https://github.com/dfm/python-bls} \citep{BLS} at the required SNR
threshold. Figure \ref{fig:folded} shows the injected transit in the
lightcurve, folded on the period of the injected planet (grey
points). Also shown is the re-binned lightcurve after folding on the period of the planet (blue points) and the model for the
injected transit (red line). 

This is of particular importance since the current sample of
    known transiting planets around evolved hosts in the NASA
    Exoplanet Archive all have a detection SNR$\geq 15$. Returning
    once more to Kepler-56, the detection ratios in that system are 63
    and 44, for planets b and c respectively. However as the BLS
    injection test shows, smaller planets are recoverable in the data.
    The transit injection performed here, along with the minimum
    planet radii calculated above, suggest that Neptune-sized planets
    should be detectable in the \emph{Kepler} lightcurves of
    low-luminosity, red-giant stars, if they are present.

\section{Conclusions}

In this work we have presented a simple model to describe the noise
properties of evolved stars as relevant to transit searches for
exoplanets.  Our model predictions of the commonly-used \emph{Kepler}
CDPP noise metric is dominated for evolved stars by granulation and
oscillations. It includes a significant contribution from stellar
oscillations, with the solar-like oscillations representing the
dominant noise source for any photometric survey of stars near the tip
of the red-giant branch. Importantly, our model also recovers the
appropriate noise signatures for highly evolved stars, a feature not
shared by current \emph{Kepler} results. This noise model may be
applied to the predictions of the noise properties of evolved stars
for the upcoming TESS and PLATO missions.

As a simple application of this updated CDPP, we also
    estimated minimum detectable planet radii for low-luminosity red
    giants, for different assumed orbital periods.  The results
    suggest that Neptune-sized planets on short-period ($P \le 20\,\rm
    days$) orbits should be detectable in the \emph{Kepler} data.  We
    advocate a detailed search for planets around red giants. Giant
    planets around evolved stars will also be detectable in
    lightcurves from the upcoming TESS mission \citep{TESS} as well as
    the ongoing K2 mission, which has already targeted a dedicated
    sample of several thousand low-luminosity red giants to detect
    giant planets (\citealt{2015H}, Grunblatt et al., submitted).

\section*{Acknowledgements}

Funding for this Discovery mission is provided by NASA's Science
Mission Directorate. The authors wish to thank the entire Kepler team,
without whom these results would not be possible. T.S.H.N., W.J.C. and
T.L.C. acknowledge financial support from the UK Science and Technology
Facilities Council (STFC). Funding for the Stellar Astrophysics Centre
is provided by The Danish National Research Foundation (Grant
agreement no.: DNRF106), M.N.L. acknowledges the support of The Danish Council for Independent Research | Natural Science (Grant DFF-4181-00415).
The research is supported by the ASTERISK project (ASTERoseismic Investigations with SONG and Kepler) funded by the European Research Council (Grant agreement no.:
267864). 

T.S.H.N. would also like to acknowledge Yvonne Elsworth for the identification of \emph{Kepler} RGB stars.
D.H. acknowledges support by the Australian Research Council's Discovery Projects funding scheme (project number DE140101364) and support by the National Aeronautics and Space Administration under Grant NNX14AB92G issued through the \emph{Kepler} Participating Scientist Program.



\bibliographystyle{mnras}
\bibliography{biblio.bib} 


\bsp	
\label{lastpage}
\end{document}